\documentclass[epj, twocolumn]{webofc}
\usepackage[varg]{txfonts}   

\woctitle{LHCP 2013}

\RequirePackage{graphicx}
\RequirePackage{mathptmx}      
\RequirePackage{flushend}
\RequirePackage[colorlinks,citecolor=blue,urlcolor=blue,linkcolor=blue]{hyperref}
\usepackage{amsmath}

\begin{document}

\title{New Physics searches in heavy flavours in ATLAS}
%
%

\author{Stefanos Leontsinis\inst{1,2}\fnsep\thanks{\email{Stefanos.Leontsinis@cern.ch}} \\ on behalf of the ATLAS Collaboration 
}

\institute{National Technical University of Athens 
\and
           Brookhaven National Laboratory
          }

\abstract{
Precision determinations of the flavour sector allow the search for indirect new physics signatures. 
At the forefront of these studies are the determinations of interference of new physics with known $\upDelta F=1$ and $\upDelta F=2$ processes. The ATLAS collaboration explores this area with competitive results measuring the CP violating phase $\phi_s$ from $B^0_s\to J/\psi \phi$ decays and investigating rare $B$ decays with dileptons in the final state with data collected at the Large Hadron Collider. In this paper, the latest ATLAS results relevant for new physics searches in the heavy flavour sector will be discussed.
}

\maketitle

\section{Introduction}\label{intro}
The ATLAS detector\cite{RefATLAS} is a general purpose detector located at the Large Hadron Collider (LHC). ATLAS has a dedicated $B$-physics program that concentrates on low momentum di-muon signatures which can efficiently be triggered.
This article focuses on the measurement of $B_s^0\to J/\psi\phi$ decay parameters\cite{RefJpsiphi}, the measurement of the forward-backward asymmetry ($A_{\mathrm{FB}}$) and the fraction of the $K^{* 0}$ longitudinal polarization ($F_\mathrm{L}$) in the decay $B^0_d\to K^{* 0}\mu^+\mu^-$\cite{RefK0starmumu} and the search for rare decays studying $B_s^0\to\mu^+\mu^-$ \cite{RefBsmumu}.

\subsection{The ATLAS detector}\label{sec:atlas_detector}
The ATLAS detector is described in detail in reference \cite{RefATLAS}. The sub-detectors of greatest importance for $B$ physics analyses are the inner detector (ID) and the muon spectrometer (MS). Both systems are surrounded by a large magnetic field to determine the momentum of the charged tracks with high precision. The mass resolution is between $\sigma=46-111\, \mathrm{MeV}$, broadening with increasing rapidity. The ID covers a pseudorapidity ($\eta$) range up to $|\eta|=2.5$ and the MS up to $|\eta|=2.7$. The tracking devices consist of high precision silicon detectors, and they are used to reconstruct the production and the decay vertices of $B$-hadrons. The muon detectors are used to identify and to trigger on muon decays of $B$-hadrons.

During 2011 ATLAS recorded $5.6\,\mathrm{fb^{-1}}$ of data from $pp$ collisions at a center of mass energy $\sqrt{s}=7\,\mathrm{TeV}$ and during 2012 recorded $21.7\,\mathrm{fb^{-1}}$ at $\sqrt{s}=8\,\mathrm{TeV}$.

\subsection{Search for new physics}\label{sec:Search_for_New_Physics}
Although Standard Model provide predictions and solutions for many physics aspects \cite{RefA}, there are persistent problems that require the presence of new physics \cite{RefB}. The heavy flavour sector is a good probe in the search of new physics. Studying leptonic and semi-leptonic decays of $b$ hadrons is an important part in the $B$-physics program of the ATLAS experiment. 

Flavour changing neutral current processes like $b\to s$ transitions in the $B_s^0\to\mu^+\mu^-$ decay and other transitions like $b\to sl^+l^-$ in the $B_s^0\to K^*\mu^{+}\mu^{-}$ decay, occurs at the lowest order via loop level diagrams within the Standard Model and have a very small branching fraction. Therefore are sensitive to effects of physics beyond the Standard Model.

The $B_s^0\to J/\psi\phi$ decay  allows the measurement of the $B_s^0$ mixing phase of the $b\to c\bar{c}s$ transition, which is responsible for CP violation in this decay mode. The Standard Model prediction for this CP violating phase is small, $\mathcal{O}(10^{-2})$, so any measured excess would be a clear indication of new physics phenomena entering in the $B_s^0$ system.

\section{CP-Violation in $B^0_s\to J/\psi\phi$}
CP violation can occur in the decay $B_s\to J/\psi\phi$ via the interference of mixing, where the $B_s^0$ oscillates into a $\bar{B}_s^0$, and the direct decay. This oscillation between the two eigenstates $(B_{H}$ and $B_{L}$) is characterized by the mass difference $(\upDelta m_s)$ and the CP-violating phase $(\phi_s)$. The light state is expected to have a shorter lifetime compared to the heavier state and the Standard Model predicts $\upDelta\Gamma_s=0.087\pm0.0021\,\mathrm{s^{-1}}$ and a very small value for $\phi_s=-0.0368\pm 0.0018\, \mathrm{rad}$.

In the $B_s^0\to J/\psi\phi$ decay, the pseudo-scalar $B_s^0$ decays to two vector mesons ($J/\psi$ and $\phi$). Due to total angular conservation the final state is an admixture of CP-odd $(L=1)$ and CP-even $(L=0,2)$ states, which can be disentangled with an angular analysis. A triplet of angular coordinates $\Omega=\left(\psi_T, \theta_T, \phi_T \right)$, defined in the transversity basis, can uniquely identify the angular signature of the decay.

For the measurement of the mixing phase $\phi_s$ of the $B^0_s\to J/\psi\phi$ decay, $4.9\,\mathrm{fb^{-1}}$ of data of $pp$ collisions with $\sqrt{s}=7\,\mathrm{TeV}$ were analyzed. ATLAS has updated its untagged analysis of the $B_s^0\to J/\psi\phi$ decay published in 2012\cite{RefoldJpsiphi} applying flavour tagging.

\begin{figure}[h]
\begin{minipage}{\columnwidth}
\centering
  \includegraphics[width=0.49\textwidth]{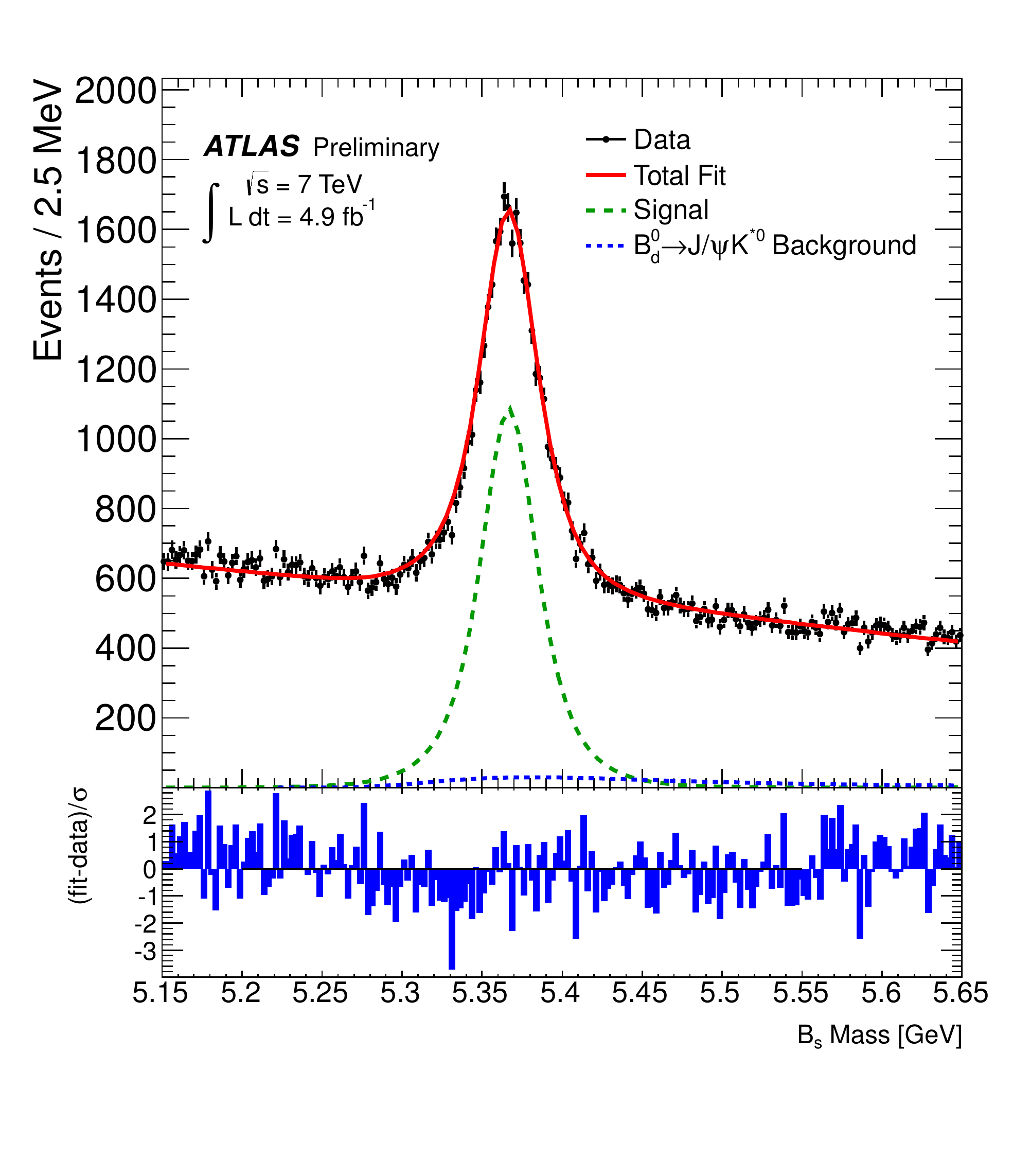}
  \includegraphics[width=0.49\textwidth]{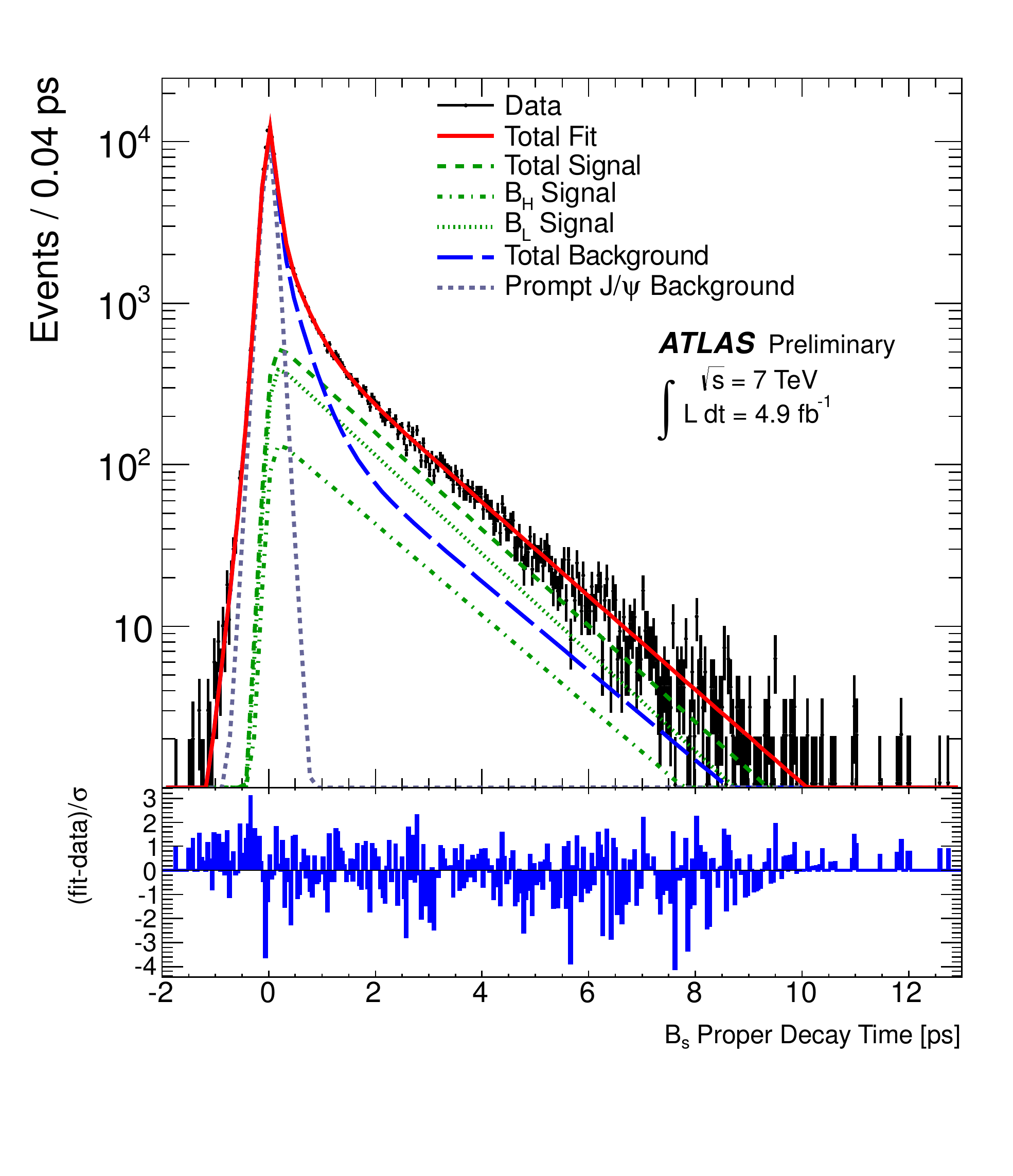}\\
  \includegraphics[width=0.49\textwidth]{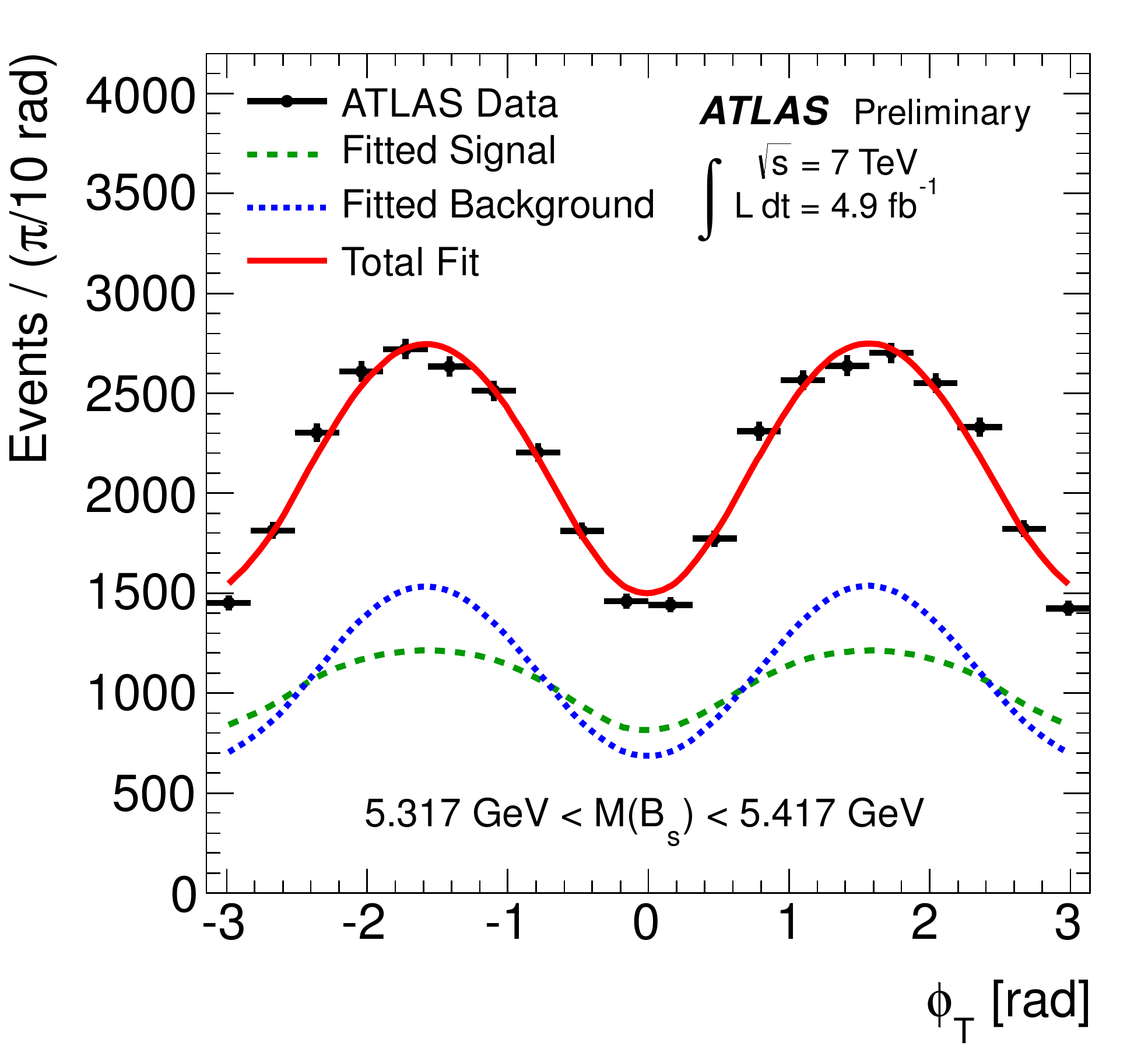}
  \includegraphics[width=0.49\textwidth]{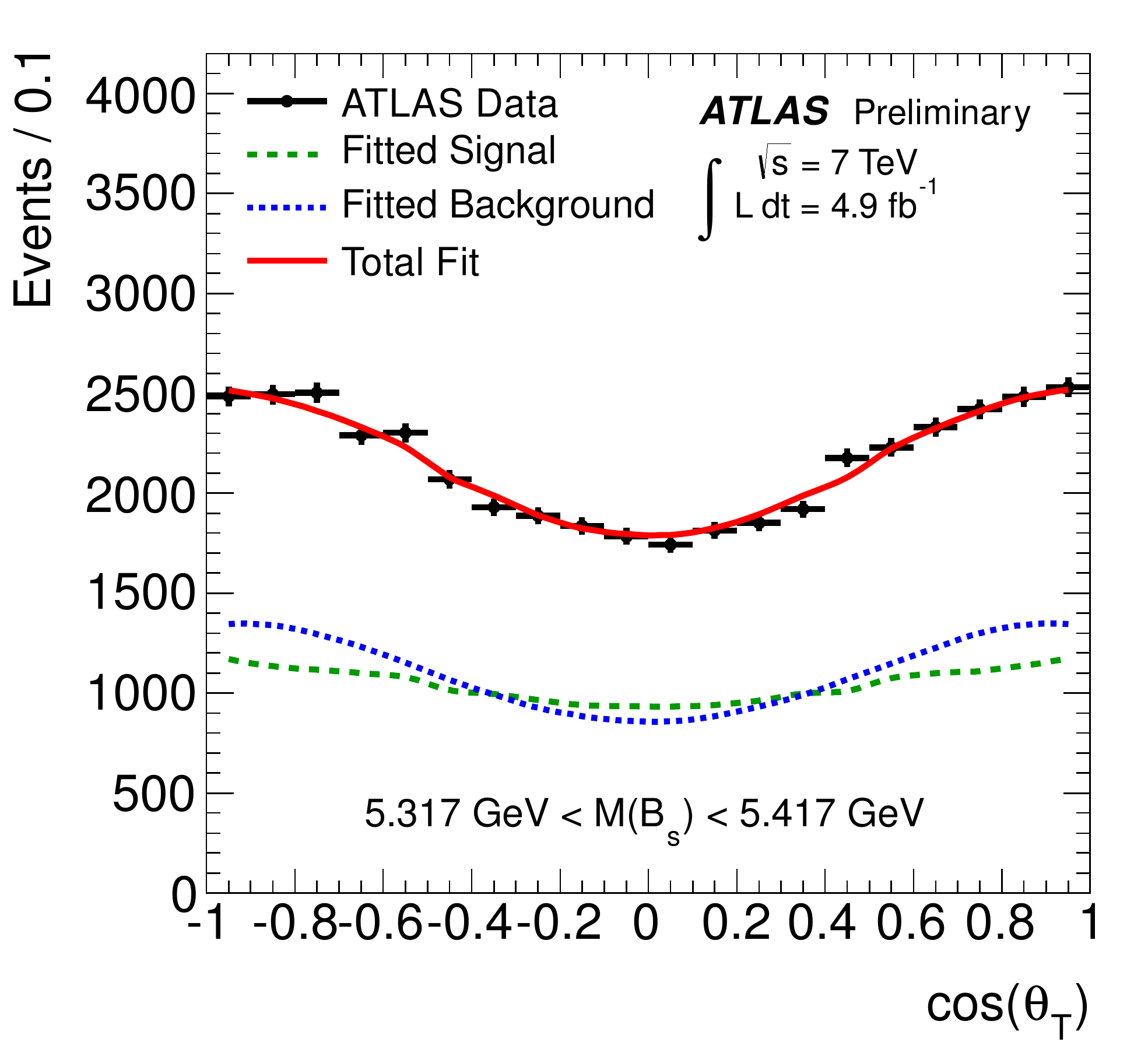}\\
  \includegraphics[width=0.49\textwidth]{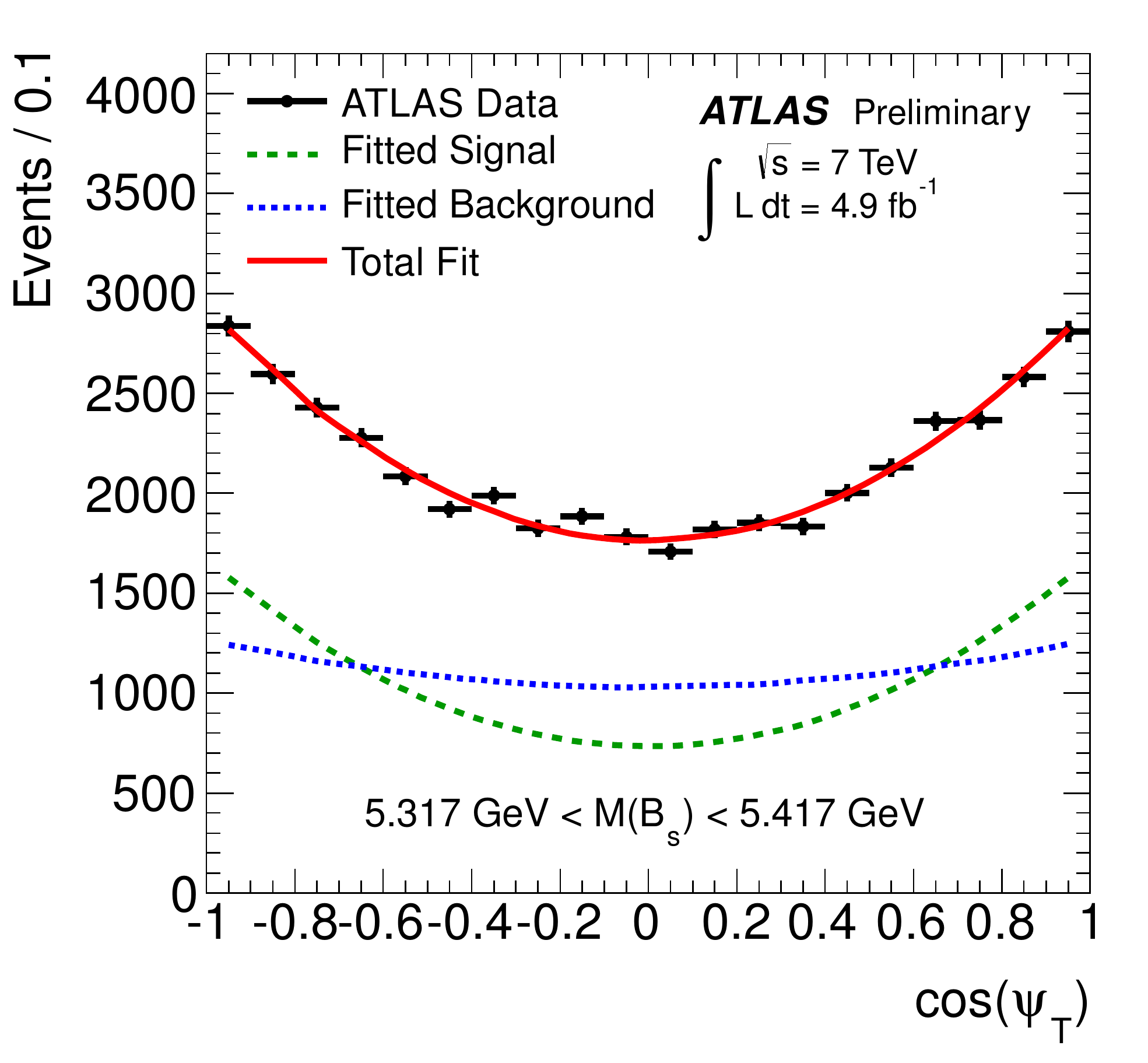}
\caption{\textit{Top:} (Left) Mass fit projection for the $B_s^0$. (Right) Proper decay time fit projection for the $B_s^0$. The pull distributions at the bottom show the difference between data and fit value normalized to the data uncertainty.
\textit{Bottom:} Fit projections for transversity angles. (Left): $\phi_T$, (Right): $\cos\theta_T$, (Center): $\cos\psi_T$\cite{RefJpsiphi}.}
\end{minipage}
\label{fig:Jpsiphifitprojections}       
\end{figure}

The measurement is based on a five-dimensional, unbinned maximum likelihood simultaneous fit to mass, proper-time and angular coordinates $\left(\psi_T, \theta_T, \phi_T \right)$. The fit model contains $25$ free parameters, nine of them being the physics parameters of interest: $(\mathrm{\upDelta\Gamma}_s)$, $\phi_s$, $\mathrm{\Gamma}_s$, $|A_0(0)|^2$, $|A_{||}(0)|^2$, $\delta_{||}$, $\delta_\perp$, $|A_s|^2$ and $\delta_s$. Results of this fit can be seen in Figure~\ref{fig:Jpsiphifitprojections}, through the projections of the distributions of mass, proper decay time and the three transversity angles.

Flavour tagging information enters into the fit model in form of a probability that the $B$ candidate is either a $B_s^0$ or a $\bar{B}_s^0$. ATLAS uses two opposite-side flavour taggers, both using $B^\pm\to J/\psi K^\pm$ decays in data for calibration and performance studies. The performance of the tagging algorithms is shown in Table~\ref{jpsiphitable}.

The first tagger is called the muon cone charge-tagger and uses muons from the semi-leptonic $B$ decay of the non-signal candidate in the event having originated near the original interaction point. Muons are separated into two reconstruction classes: 1) "Combined muons", when the muon is reconstructed using information form both the ID and the MS and 2) "Segment Tagged", when the muon is formed by segments which are not associated with an MS track, but which are matched to ID tracks extrapolated to the MS. A muon cone charge variable is calculated from tracks found in the inner tracking detectors that are in a cone of $\upDelta R<0.5$ around the momentum axis of the muon. 

The second tagging method is the jet charge tagger. In the absence of a muon, the events are searched for a b-tagged jet, with tracks associated to the same primary vertex as the signal decay, excluding those from the signal candidate. The jet is reconstructed using the anti-$k_t$ algorithm with a cone size of $0.6$.

The muon cone and the jet charge is defined as:
\begin{equation*}
Q_{\mu,\mathrm{jet}}=\frac{\sum_i^{\mathrm{N_{tracks}}} q^i\cdot (p_{\mathrm{T}}^i)^\kappa}{\sum_i^{\mathrm{N_{tracks}}} (p_{\mathrm{T}}^i)^\kappa}
\end{equation*}

where $\kappa=1.1$, which was tuned to optimize the tagging power.

\begin{table*}[t]
\centering
\caption{Table showing the performance of the b-tagger}
\label{jpsiphitable}
\begin{tabular}{c c c c}
\hline
Tagger & Efficiency [\%] & Dilution [\%] & Tagging Power [\%]\\
\hline
Segment Tagged  muon           & $1.08\pm 0.02$     & $36.7\pm 0.7$        & $0.15\pm 0.02$          \\
Combined muon & $3.37\pm 0.04$ & $50.6\pm 0.5$ & $0.86 \pm 0.04$ \\
Jet charge & $27.7\pm 0.1$ & $12.68\pm 0.06$ & $0.45\pm 0.05$\\
Total & $32.1\pm 0.1$ & $21.3\pm 0.08$ & $1.45\pm 0.05$\\
\hline
\end{tabular}
\end{table*}

The measured values of the physics parameters are:

\begin{eqnarray*}
  \phi_s & = & 0.12\pm0.25\mathrm{(stat.)}\pm 0.11\mathrm{(syst.)}\,\mathrm{rad}\\
  \mathrm{\Delta}\Gamma_s & = & 0.058\pm 0.021\mathrm{(stat.)}\pm 0.009\mathrm{(syst.)}\,\mathrm{ps^{-1}}\\
  \Gamma_s & = & 0.677\pm 0.007\mathrm{(stat.)}\pm 0.003\mathrm{(syst.)}\,\mathrm{ps^{-1}}\\
  |A_0(0)|^2 & = & 0.529\pm 0.006\mathrm{(stat.)}\pm 0.011\mathrm{(syst.)}\\
  |A_{\|}(0)|^2 & = & 0.220\pm 0.008\mathrm{(stat.)}\pm 0.009\mathrm{(syst.)}\\
  |A_s(0)|^2 & = & 0.024\pm 0.014\mathrm{(stat.)}\pm 0.028\mathrm{(syst.)}\\
  \delta_{\perp} & = & 3.89\pm 0.46\mathrm{(stat.)}\pm 0.13\mathrm{(syst.)}\,\mathrm{rad}
\end{eqnarray*}

\begin{figure}[h!]
\begin{minipage}{\columnwidth}
\centering
  \includegraphics[width=1\textwidth]{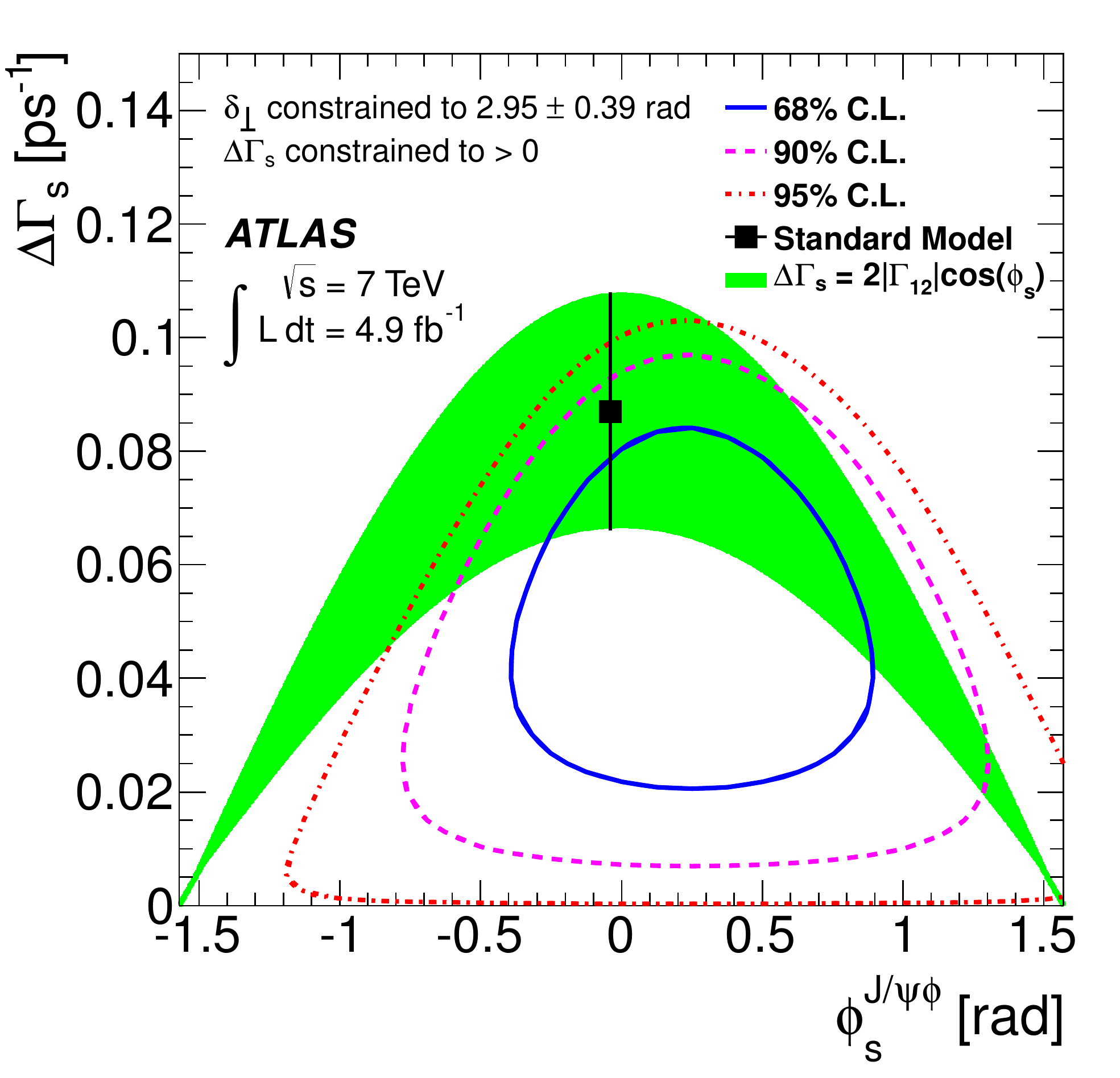}
  \includegraphics[width=1\textwidth]{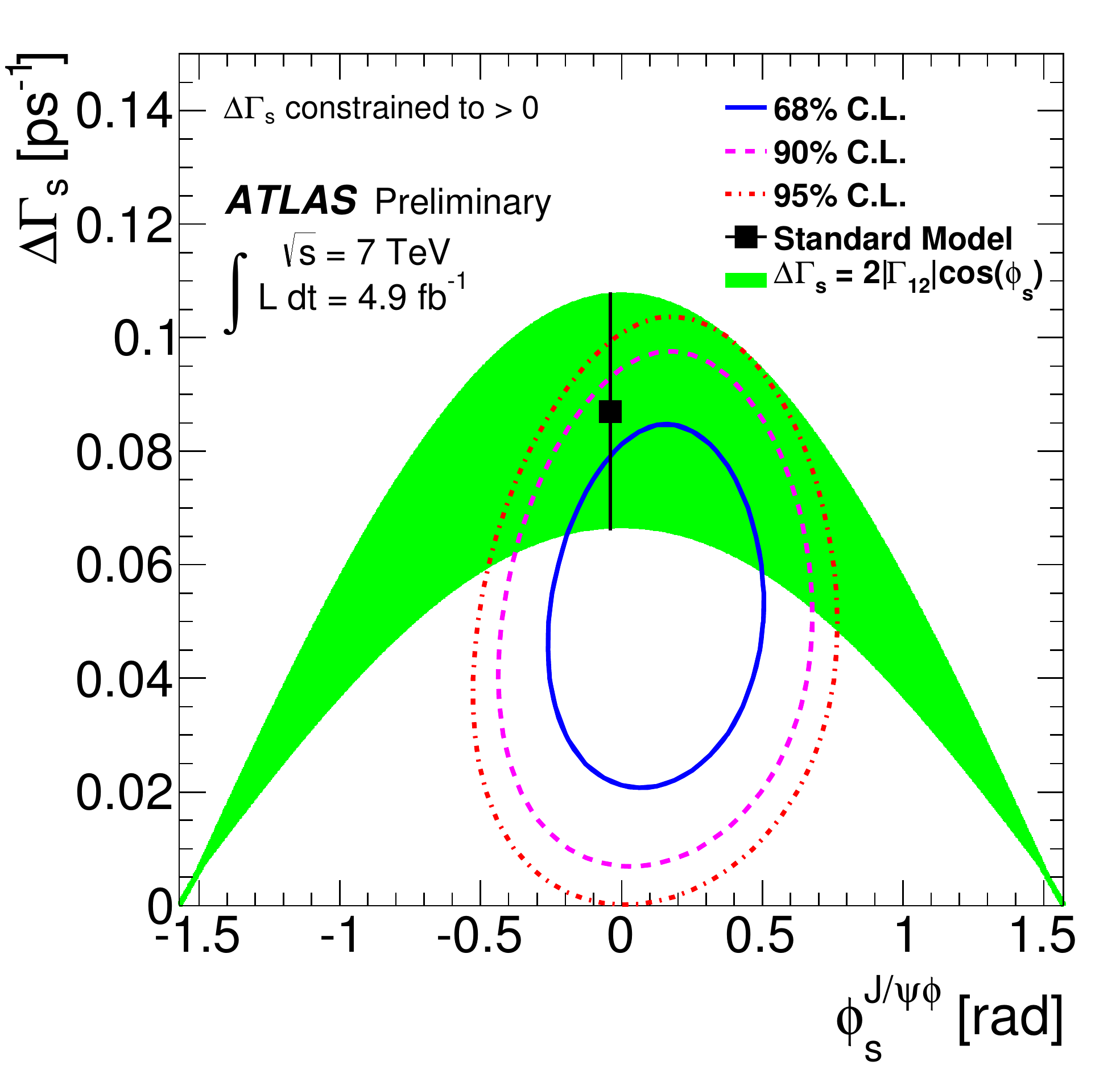}
\caption{Likelihood contours in $\phi_s-\upDelta\Gamma_s$ plane. The blue and red contours show the 68\% and 95\% likelihood contours, respectively (statistical errors only). The green band is the theoretical prediction of mixing-induced CP violation. \textit{Top:} Previous ATLAS measurement, without using $B$ flavour tagging\cite{RefoldJpsiphi}, \textit{Bottom:} New result including flavour tagging\cite{RefJpsiphi}.}
\end{minipage}
\label{fig:Jpsiphiresult}
\end{figure}

Concluding on the $B_s^0\to J/\psi\phi$, the obtained values of the parameters are in agreement with the Standard Model and other experimental measurements\cite{CDFJpsiphi}\cite{LHCbJpsiphi}. Using the flavour tagging information resulted in a $\sim 40\%$ improvement in the uncertainty of $\phi_s$ compared to the previous analysis\cite{RefoldJpsiphi}. The likelihood contour plot in the $\phi_s - \mathrm{\Delta}\Gamma_s$ plane for the tagged and un-tagged analyses are shown in Figure~\ref{fig:Jpsiphiresult}.

\section{$B^0_{d}\to{K^{* 0}}\mu^+\mu^-$ Forward-Backward Asymmetry}

Within the Standard Model the decay $B^0_{d}\to{K^{* 0}}\mu^+\mu^-\to K^+\pi^-\mu^+\mu^-$  has a small branching fraction of $(1.06\pm 0.1)\times 10^{-6}$\cite{RefPDG}. The angular distributions of the four-particle final state and the decay amplitudes are sensitive to physics beyond the Standard Model, due to the interference of new diagrams with the Standard Model ones.

\begin{figure}[h]
  \centering
  \includegraphics[width=.4\textwidth]{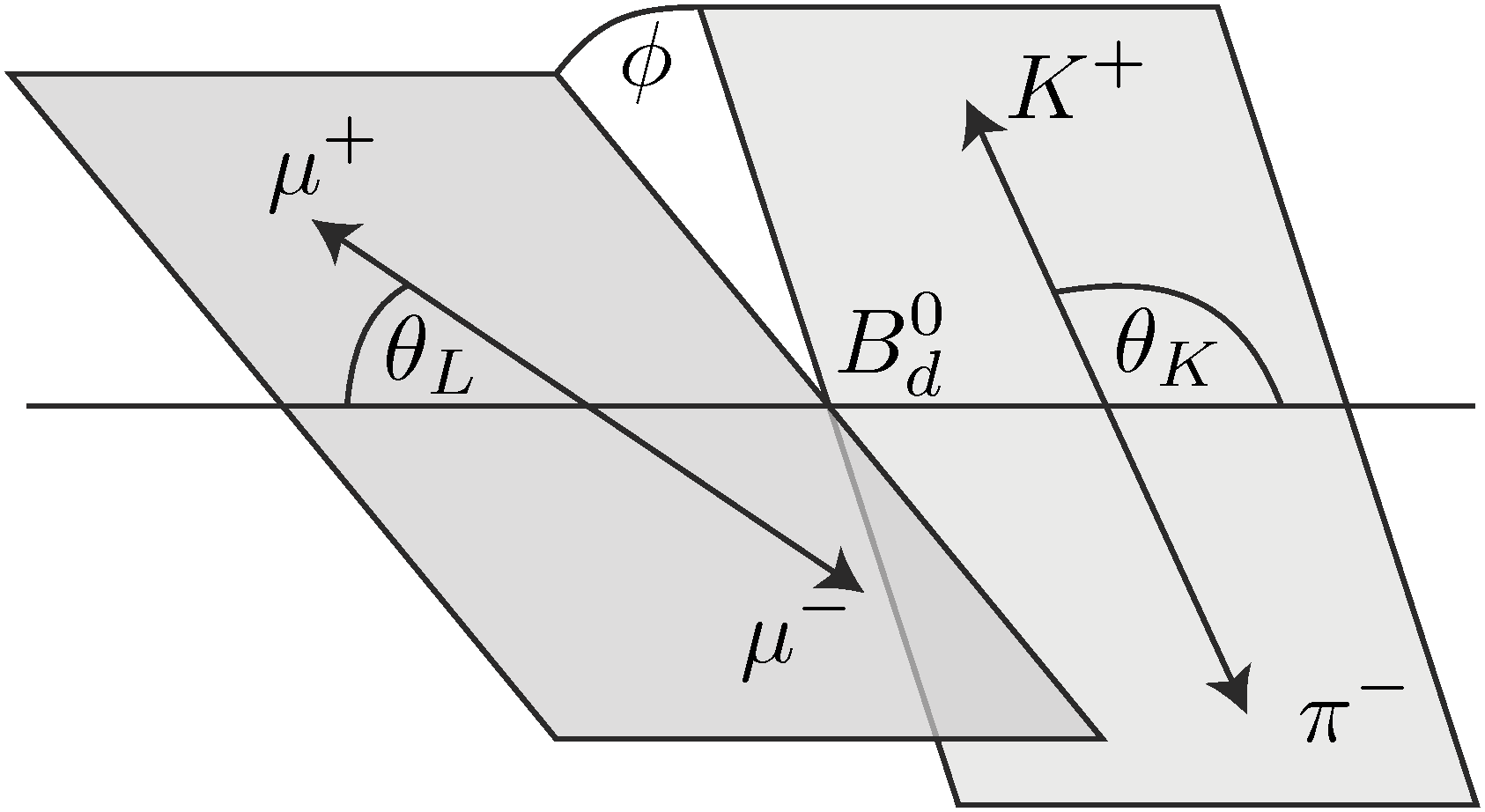}
  \caption{Definition of the kinematic angles in the decay $B_d^0\to K^{* 0}\mu^+\mu^-$\cite{RefK0starmumu}.}
\label{fig:K0starangles}       
\end{figure}

The decay of $B_d^0\to K^{* 0}\mu^+\mu^-$  is described by four kinematic variables. The invariant mass of the di-muon system $\left(q^2\right)$ and three angles describing the geometrical configuration of the final state as shown in Figure~\ref{fig:K0starangles}. $\theta_L$ is the angle between the $\mu^+$ and the direction opposite to the $B^0_d$ in the $K^{* 0}$ rest frame, $\theta_K$ is the angle between the $K^+$ and the direction opposite to the $B^0_d$ in the $K^{* 0}$ rest frame, and $\phi$ is the angle between the plane defined by the two muons and the plane defined by the kaon-pion system in the $B^0_d$ rest frame. In the case of the $\bar{B}^0_d$ the angles $\theta_L$ and $\theta_K$ are defined with respect to the $\mu^-$ and the $K^-$, respectively.

Due to limited statistics to study the four-dimensional differential decay rate, the differential decay rate is projected from the four kinematic variables into the two-dimensional distributions $\mathrm{d}^2\mathrm{\Gamma}/\mathrm{d}q^2\mathrm{d}\cos\theta_L$ and $\mathrm{d}^2\mathrm{\Gamma}/\mathrm{d}q^2\mathrm{d}\cos\theta_K$ by integrating over the two other variables. The $K^{* 0}$ longitudinal polarization fraction $(F_\mathrm{L})$ and the lepton forward-backward asymmetry $(A_{\mathrm{FB}})$ are extracted and averaged in the $q^2$ bins.

The values for $A_{\mathrm{FB}}$ and $F_\mathrm{L}$ are extracted by performing a sequential unbinned maximum likelihood fit, where in a first step the invariant $K^\pm\pi^\mp\mu^+\mu^-$ mass distribution is fitted, then the resulting parameters are fixed, and in a second step the angular distributions are fitted. An example of one $q^2$ bin of the fit can be seen in Figure~\ref{fig:K0starfit}, by projecting the fit model to $q^2$ and the two angles $(\theta_L$ and $\theta_K)$.

\begin{figure}[h!]
\centering
  \includegraphics[width=0.48\textwidth]{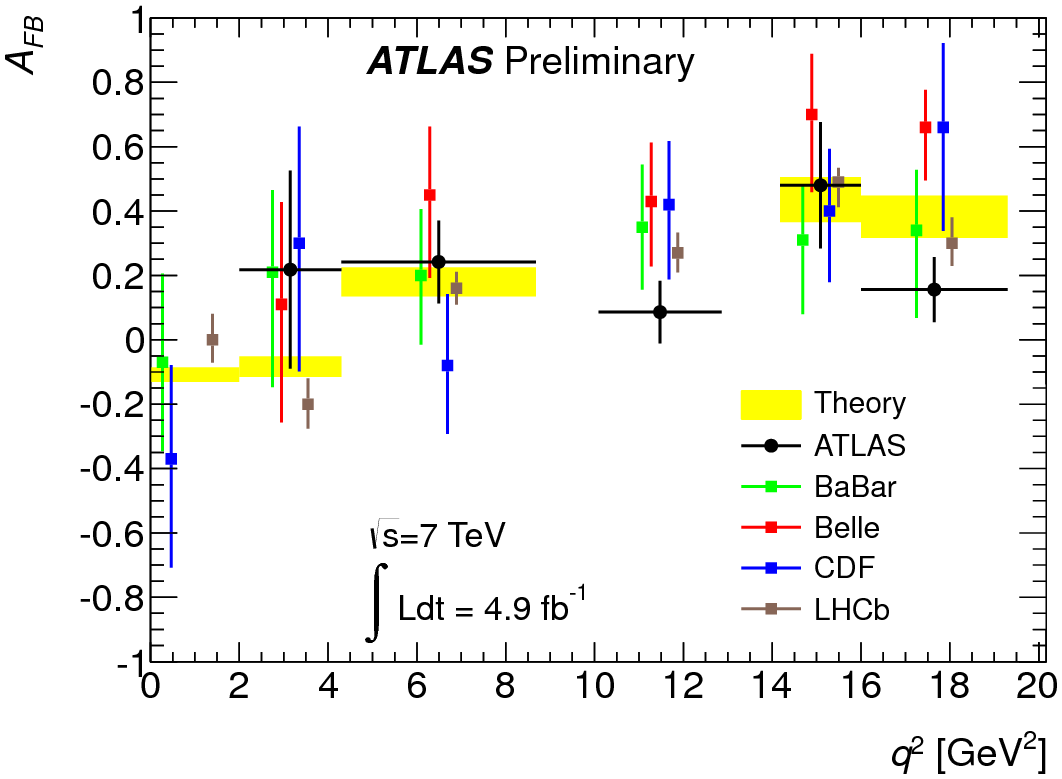}
  \includegraphics[width=0.48\textwidth]{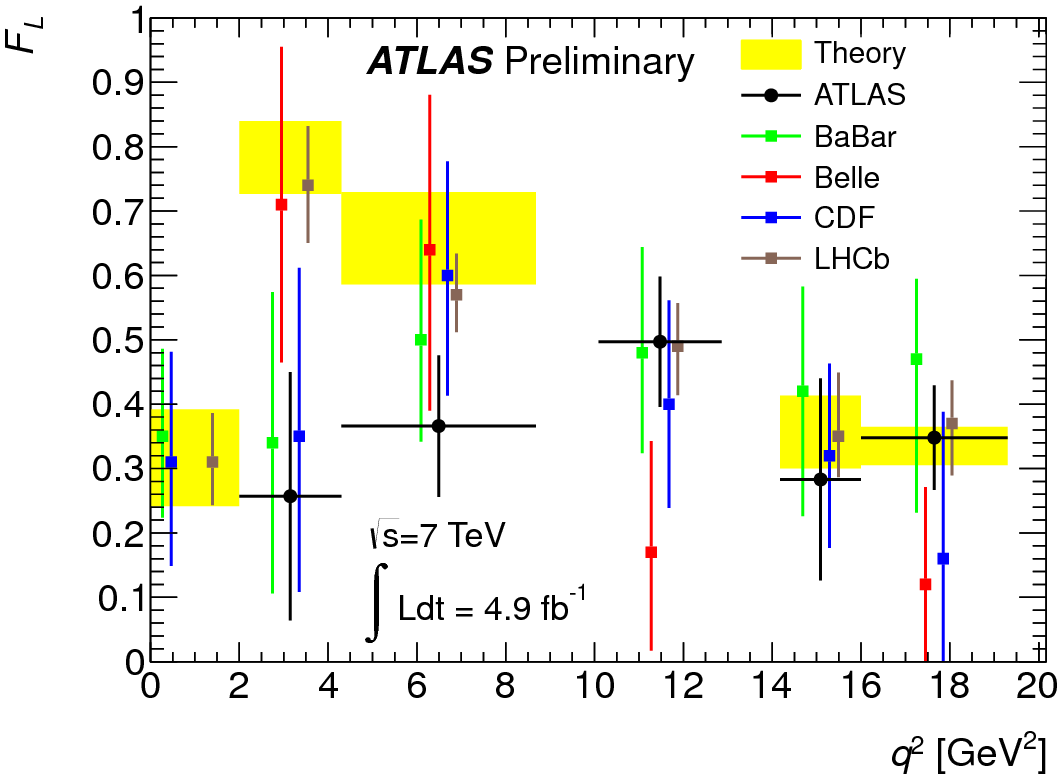}
\caption{Forward-backward asymmetry $A_{\mathrm{FB}}$ (left) and fraction of longitudinal polarized $K^{*0}$ mesons $F_\mathrm{L}$ (right) as a function of $q^2$ measured by ATLAS (black dots). In each $q^2$ bin, ordered from right to left, results of other experiments are shown as coloured squares: BaBar, Belle, CDF and LHCb. All errors including statistical and systematic uncertainties. The experimental results are compared to theoretical Standard Model predictions including theoretical uncertainties\cite{RefK0starmumu}.}
\label{fig:afbfl}       
\end{figure}

The parameters $A_{\mathrm{FB}}$ and $F_\mathrm{L}$ are extracted in five $q^2$ bins from $2\,\mathrm{GeV}^2$ to $19\,\mathrm{GeV}^2$ and in the wider bin $1\,\mathrm{GeV}^2<q^2<6\,\mathrm{GeV}^2$. The results of the unbinned maximum likelihood fit including statistical and systematic uncertainties are summarized in Table~\ref{kstartable} and are compared to Standard Model expectations and measurements from other experiments  in Figure~\ref{fig:afbfl}.

\begin{table*}[t]
\centering
\caption{Summary of the fit results for the different bins of $q^2$. number of signal events $N_{sig}$ from the mass fit and its statistical uncertainty, forward backward asymmetry $A_{\mathrm{FB}}$ and longitudinal polarization $F_\mathrm{L}$ for different bins in $q^2$ including statistical and systematic uncertainties.}
\label{kstartable}
\begin{tabular}{lclclclc|}
\hline
	$q^2$ range $\left(\mathrm{GeV}^2\right)$ & \hspace{1mm} $N_{sig}$ &\hspace{8mm} $A_{\mathrm{FB}}$ & $F_\mathrm{L}$\\
\hline
	{$\begin{aligned}[t]
		2.00<&q^2<4.30 \\
		4.30<&q^2<8.68 \\
		10.09<&q^2<12.86 \\
		14.18<&q^2<16.00 \\
		16.00<&q^2<19.00 \\
		1.00<&q^2<6.00 \\
	\end{aligned}$}
&
	{$\begin{aligned}[t]
		19&\pm 8 \\
		88&\pm 17 \\
		138&\pm 31 \\
		32&\pm 14 \\
		149 &\pm 24 \\
		42&\pm 11 \\
	\end{aligned}$}
&
	{$\begin{aligned}[t]
		0.22 &\pm 0.28 \pm0.14  \\
		0.24 &\pm 0.13 \pm 0.01\\
		0.09 &\pm 0.09 \pm 0.03 \\
		0.48 &\pm  0.19 \pm 0.05 \\
		0.16 &\pm 0.10 \pm 0.03 \\
		0.07 &\pm 0.20 \pm 0.07 \\
	\end{aligned}$}
&
	{$\begin{aligned}[t]
		0.26 &\pm 0.18 \pm 0.06  \\
		0.37 &\pm 0.11 \pm 0.02 \\
		0.50 &\pm 0.09 \pm 0.04 \\
		0.28 &\pm  0.16 \pm 0.03 \\
		0.35 &\pm 0.08 \pm 0.02 \\
		0.18 &\pm 0.15 \pm 0.03 \\
	\end{aligned}$}
\\\hline
\end{tabular}
\end{table*}

Using $4.9\,\mathrm{fb^{-1}}$ of integrated luminosity at $\sqrt{s}=7\,\mathrm{TeV}$, the forward backward asymmetry $A_{\mathrm{FB}}$ and the $K^{* 0}$ longitudinal polarization $F_\mathrm{L}$ have been measured as function of the di-muon mass squared $q^2$. The results obtained on $A_{\mathrm{FB}}$ and $F_\mathrm{L}$ are mostly consistent with theoretical predictions \cite{RefK0starTheory} and measurements performed by other experiments \cite{RefK0BaBar}, \cite{RefK0Belle}, \cite{RefK0CDF}, \cite{RefK0LHCb}.

\section{Rare Decays}
The theoretical prediction for the branching ratio of $B_s^0\to\mu^+\mu^-$ is $(3.56\pm 0.18)\times 10^{-9}$\cite{RefTheoryBsmumu}. The first analysis performed by ATLAS \cite{RefBsmumu} use $2.4\,\mathrm{fb^{-1}}$ of data collected in 2011, the total amount of data collected on a single di-muon trigger, before a change in the triggering algorithm.

The branching ratio is measured with respect to the reference channel $B^\pm\to J/\psi K^\pm \to\mu^+\mu^- K^\pm$ and computed as:

\begin{figure}[h!]
\centering
  \includegraphics[width=0.41\textwidth]{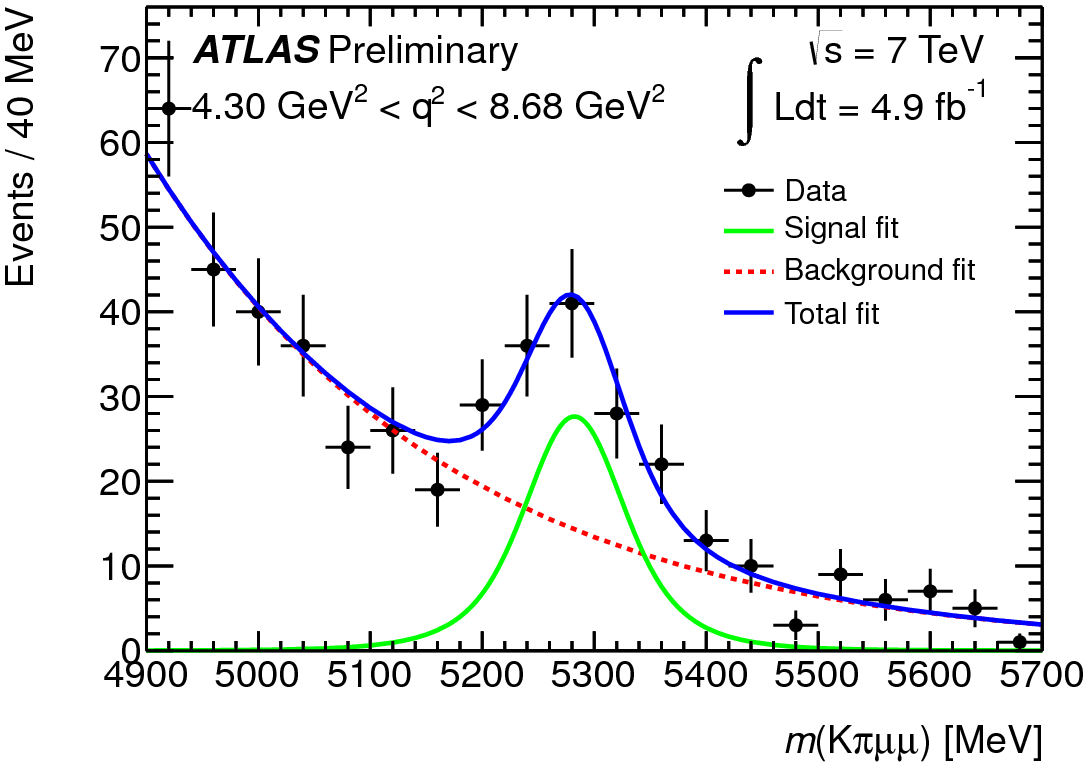}\\
  \includegraphics[width=0.23\textwidth]{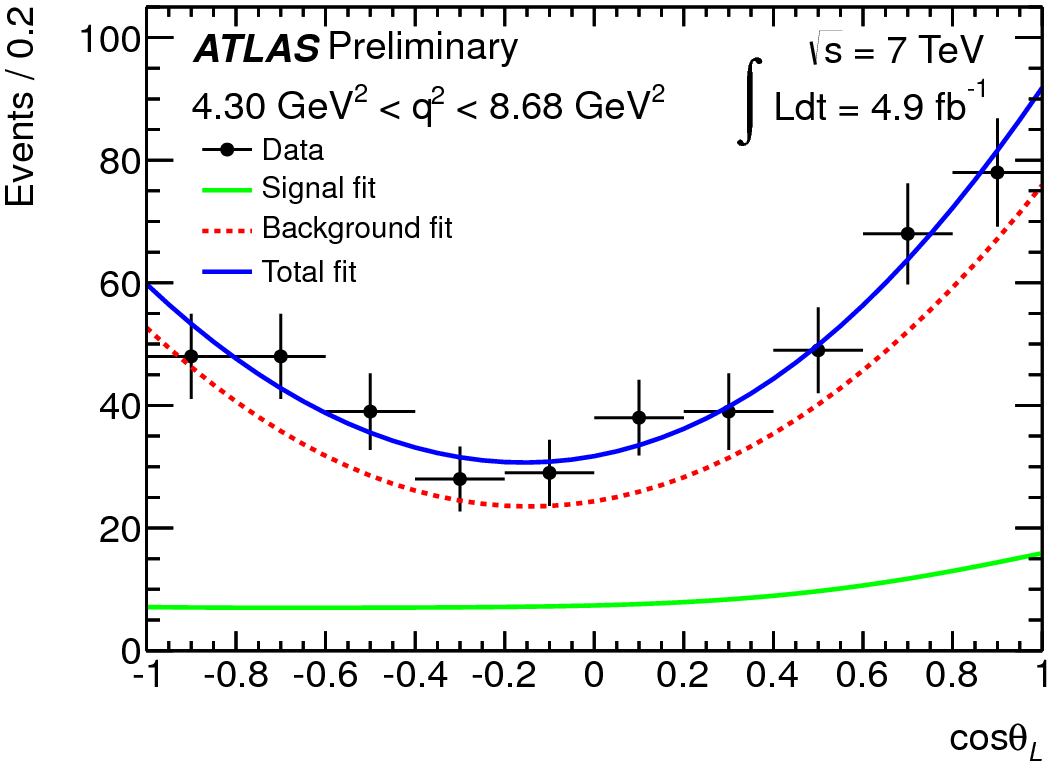}
  \includegraphics[width=0.23\textwidth]{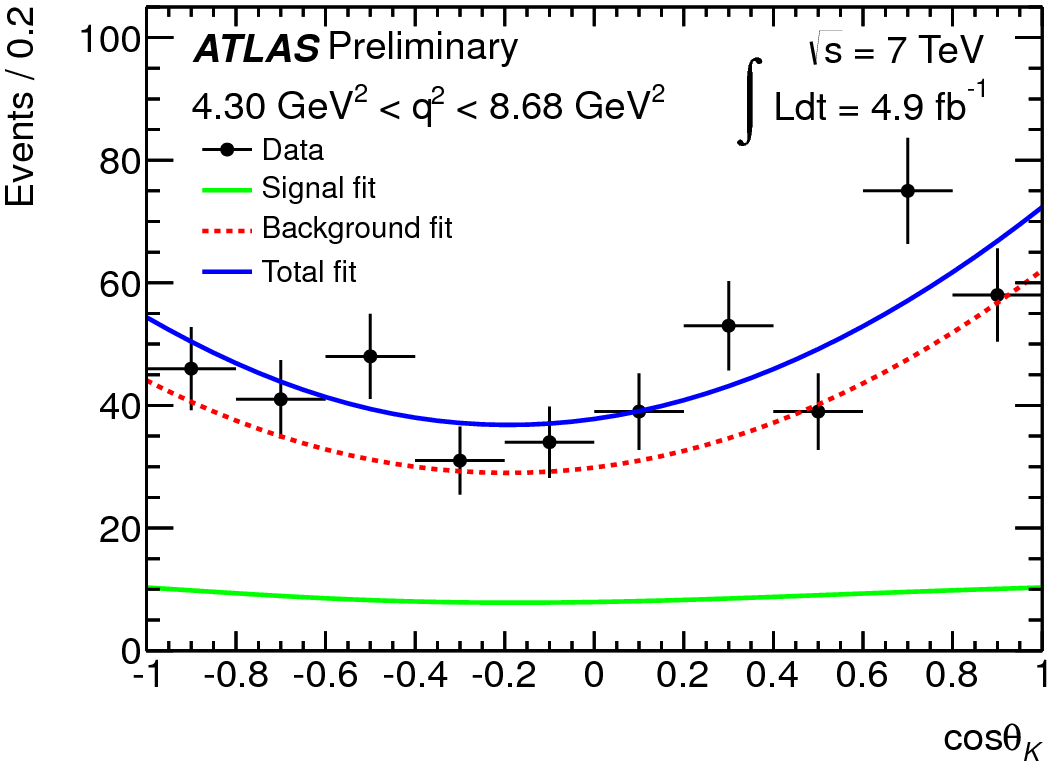}
\caption{Mass fit (top), angular fit in $\cos\theta_L$ (bottom left) and in $\cos\theta_K$ (bottom right) for the $q^2$ region between $4.3\, \mathrm{GeV^2}$ and $8.68\,\mathrm{GeV^2}$, where the data points show the statistical uncertainty, the solid blue (dark) line is the corresponding fit with the green solid line (light) the signal contribution and the red dotted line the background contribution\cite{RefK0starmumu}.}
\label{fig:K0starfit}       
\end{figure}

\begin{multline*}
{BR(B_s^0\to\mu^+\mu^-)=\frac{N_{\mu^+\mu^-}}{N_{J/\psi K^\pm}}\times \frac{\alpha_{J/\psi K^\pm}\cdot\epsilon_{J/\psi K^\pm}}{\alpha_{B_s^0\to\mu^+\mu^-}\cdot\epsilon_{B_s^0\to\mu^+\mu^-}}} \times\frac{f_u}{f_s}\\ {\times BR(B^\pm\to J/\psi K^\pm\to\mu^{+}\mu^{-} K^\pm)}
\end{multline*}

The reference channel yield $\left(N_{J/\psi K^\pm}\right)$ was determined from a binned likelihood fit to the invariant mass distribution of the $\mu^+\mu^-K^\pm$ system and the acceptance and efficiency ratios are estimated from MC samples. The relative production rates for $B^\pm$ and $B^0_s$ $f_s/f_u$ is $0.267\pm 0.021$\cite{Reffsfu} (assuming $f_s=f_d$ and no kinematic dependance of $f_s/f_u$). The branching fraction of the reference channel $B^\pm\to J/\psi K^\pm\to\mu^{+}\mu^{-} K^\pm$ is taken from previous measurements\cite{RefPDG} and is equal to $\left(6.01\pm 0.21 \right)\times 10^{-5}$.

The expected limit is calculated to be $2.3^{+1.0}_{-0.5}\times 10^{-8}$ at $95\%$ CL and the corresponding observed limit $2.2\times 10^{-8}$ at $95\%$ CL. LHCb collaboration recently reported first evidence for the $B_s^0\to\mu^+\mu^-$ decay with $\mathcal{B}(B_s^0\to\mu^+\mu^-)=(3.2^{+1.5}_{-1.2})\times 10^{-9}$ \cite{LHCbBs}.

\section{Conclusion}
ATLAS has a rich $B$-physics program which includes indirect searches for new physics, like the rare decay $B^0_s\to\mu^+\mu^-$, measurements of CP-violating phase in the $B_s^0\to J/\psi\phi$ decay and measurement of the forward backward asymmetry $A_{\mathrm{FB}}$ and the $K^{* 0}$ longitudinal polarization $F_\mathrm{L}$. An upper limit on the branching fraction of $B^0_s\to\mu^+\mu^-$ of $2.2\times 10^{-8}$ at $95\%$ CL has been set using $2.4\,\mathrm{fb^{-1}}$ of 2011 data. Several parameters describing the $B_s^0$ meson were measured, with $4.9\,\mathrm{fb^{-1}}$ of 2011 data, using flavour tagging and their values are consistent with the theoretical expectations. The CP-violating phase $\phi_s$ for the $B_s^0$ meson was measured to be $\phi_s=0.12\pm0.25(\mathrm{stat.})\pm0.11(\mathrm{syst.})\, \mathrm{rad}$. All results are consistent with other measurements and Standard Model predictions. Updated analyses including higher integrated luminosity and new analysis methods are expected to lower the systematic uncertainties and improve the precision of the results that now are dominated by statistical uncertainties.

\section*{Acknowledgements}
This research has been co-financed by the European Union (European Social Fund - ESF) and Greek national funds through the Operational Program "Education and Lifelong Learning" of the National Strategic Reference Framework (NSRF) - Research Funding Program: \textbf{THALES}. Investing in knowledge society through the European Social Fund.


\begin{thebibliography}{}
\bibitem{RefATLAS}
ATLAS Collaboration, The ATLAS Experiment at the CERN Large Hadron Collider, JINST, 3, S08003 (2008)
\bibitem{RefJpsiphi}
ATLAS Collaboration, Flavour tagged time dependent angular analysis of the $B^0_s\to J/\psi\phi$ decay and extraction of $\upDelta\Gamma_s$ and the weak phase $\phi_s$ in ATLAS, ATLAS-CONF-2013-039, https://cds.cern.ch/record/1541823 (2013)
\bibitem{RefK0starmumu}
ATLAS Collaboration, Angular analysis of $B^0_d\to K^{* 0}\mu^+\mu^-$ with the ATLAS Experiment, ATLAS-CONF-2013-038, https://cds.cern.ch/record/1537961 (2013)
\bibitem{RefBsmumu}
ATLAS Collaboration, Search for the decay $B^0_s\to\mu^+\mu^-$, Phys. Lett. B713, 180-196 (2012) 
\bibitem{RefA}
F. Bezrukov,  G. Karananas, J. Rubio, M. Shaposhnikov, Higgs - Dilaton Cosmology: an effective field theory approach , Phys. Rev. D, 87, 096001 (2013)
\bibitem{RefB}
M. Shaposhnikov, Is there a new physics between electroweak and Planck scales?, arXiv:0708.3550 (2007)
\bibitem{CDFJpsiphi}
T. Aaltonen et al (CDF Collaboration), Measurement of the CP-Violating Phase $\beta_s^{J/\psi\phi}$ in $B_s^0\to J/\psi\phi$, Phys. Rev. Lett., 108, 101803, arXiv:1112.3183 (2012) 
\bibitem{LHCbJpsiphi}
R. Aaij et al. (LHCb Collaboration), Measurement of the CP-violating phase $\phi_s$ in the decay $B_s^0\to J/\psi\phi$, Phys. Rev. Lett., 108, 101803, arXiv:1112.3183 (2012)
\bibitem{RefoldJpsiphi}
ATLAS Collaboration, $\phi_s$ and $\mathrm{\Delta}\Gamma_s$ from time dependent angular analysis of $B^0_s\to J\psi\phi$, JHEP, 12, 072 (2012)
\bibitem{RefPDG}
Particle Data Group, J. Beringer et al., Review of Particle Physics, Phys. Rev. D86, 010001 (2012)
\bibitem{RefK0starTheory}
C. Bobeth, G. Hiller and D. van Dyk, 
Angular analysis of $\bar{B}\to \bar{K}^{(*)} l^+ l^-$ decays, Phys.Rev. D87, 034016, arXiv:1212.2321 (2011) 
\bibitem{RefK0BaBar}
B. Aubert et al. (BaBar Collaboration), Angular Distributions in the Decays $B\to K^* l^+l^-$, Phys. Rev. D 79, 031102 (2009)
\bibitem{RefK0Belle}
J.-T. Wei, P. Chang, et al (The Belle Collaboration), Measurement of the Differential Branching Fraction and Forward-Backward Asymmetry for $B\to K^* l^+l^-$,  Phys. Rev. Lett. 103, 171801 (2009)
\bibitem{RefK0CDF}
T. Aaltonen et al (CDF Collaboration), Measurements of the Angular Distributions in the Decays $B\to K^* \mu^+ \mu^-$ at CDF,  Phys. Rev. Lett. 108, 081807 (2012)
\bibitem{RefK0LHCb}
A.A. Alves Jr. et al. (LHCb Collaboration), Differential branching fraction and angular analysis of the $B^{0}\to K^{* 0}\mu^+\mu^-$ decay, LHCb-CONF-2012-008. http://cds.cern.ch/record/1427691 (2012)
\bibitem{RefTheoryBsmumu}
A. J. Buras, R. Fleischer, J. Girrbach, and R. Knegjens, arXiv:1303.3820 [hep-ph] (2013)
\bibitem{Reffsfu}
A.A. Alves Jr. et al. (LHCb Collaboration), Measurement of $b$ hadron production fractions in $7\,\mathrm{TeV}$ $pp$ collisions, Phys. Rev. D85, 032008 (2012) 
\bibitem{LHCbBs}
R. Aaij et al. (LHCb Collaboration), First evidence for the decay $B_s\to\mu^+\mu^-$, Phys. Rev. Lett., 110, 021801, arXiv:1211.2674 (2013)
\end{thebibliography}
\end{document}